\documentclass[12pt]{article}
\usepackage{amssymb,amsmath,epsfig}
\allowdisplaybreaks

\begin{document}

\title{\bf Cosmic Evolution of Holographic Dark Energy in $f(\mathcal{G},T)$ Gravity}
\author{M. Sharif \thanks {msharif.math@pu.edu.pk} and Ayesha Ikram
\thanks{ayeshamaths91@gmail.com}\\
$^{1}$ Department of Mathematics, University of the Punjab,\\
Quaid-e-Azam Campus, Lahore-54590, Pakistan.}

\date{}
\maketitle

\begin{abstract}
The aim of this paper is to analyze the cosmological evolution of
holographic dark energy in $f(\mathcal{G},T)$ gravity ($\mathcal{G}$
and $T$ represent the Gauss-Bonnet invariant and trace of the
energy-momentum tensor, respectively). We reconstruct
$f(\mathcal{G},T)$ model through correspondence scheme using
power-law form of the scale factor. The qualitative analysis of the
derived model is investigated with the help of evolutionary
trajectories of equation of state, deceleration as well as
state-finder diagnostic parameters and
$\omega_{_{\mathcal{G}T}}-{\omega}'_{_{\mathcal{G}T}}$ cosmological
phase plane. It is found that the equation of state parameter
represents phantom epoch of the Universe whereas the deceleration
parameter illustrates the accelerated phase. The state-finder plane
corresponds to Chaplygin gas model while the freezing region is
attained in $\omega_{_{\mathcal{G}T}}-{\omega}'_{_{\mathcal{G}T}}$
plane.
\end{abstract}

\section{Introduction}

The surprising discovery of the accelerated expansion of the
Universe is one of the exciting progress in cosmology. This
tremendous change in cosmic history has been proved from a diverse
set of high-precision observational data accumulated from various
astronomical sources. The accelerating paradigm is considered as a
consequence of an exotic type of force dubbed as dark energy (DE)
which possesses repulsive characteristics with negatively large
pressure. It may predict the ultimate future of the Universe but its
salient features are still not known. To explore the perplexing
nature of DE, different approaches have been presented. The
cosmological constant $(\Lambda)$ is the simplest approach while
modified theories of gravity and dynamical DE models have also been
proposed in this regard. The cosmological constant suffers from
problems like fine tuning (large discrepancy between its theoretical
predicted and observed value) and coincidence between the observed
vacuum energy and the current matter density. Modified gravitational
theories act as an alternative for dark energy (DE) and are obtained
by replacing or adding curvature invariants as well as their
corresponding generic functions in the geometric part of the
Einstein-Hilbert action. It is found that the negative powers of
scalar curvature $(R)$ in $f(R)$ theory act as an alternative to DE
and thus produce acceleration in the cosmic expansion while its
positive powers elegantly describe the inflationary era \cite{a}.
Various modified theories possess quite rich cosmological structure,
pass the solar system constraints, efficiently describe the bouncing
cosmology as well as provide a gravitational alternative for a
unified description of the inflationary epoch to the late-time
accelerated expansion \cite{b}-\cite{b3}.

Gauss-Bonnet (GB) invariant being a particular linear combination of
quadratic curvature invariants has gained much attention in
cosmology. This four-dimensional topological invariant is free from
spin-2 ghost instabilities and is defined as \cite{1}-\cite{1b}
\begin{equation}\label{GB}
\mathcal{G}=R^{2}-4R_{\xi\eta}R^{\xi\eta}+R_{\xi\eta\mu\nu}R^{\xi\eta\mu\nu},
\end{equation}
where $R_{\xi\eta}$ and $R_{\xi\eta\mu\nu}$ are the Ricci and
Riemann tensors, respectively. To investigate the dynamics of
$\mathcal{G}$ in four dimensions, Nojiri et al. \cite{c} coupled the
GB invariant with scalar field and demonstrated that the cosmic
accelerated expansion may be produced by the mixture of scalar
phantom and/or potential/stringy effects while this scalar GB
coupling acts against the big-rip occurrence in phantom cosmology.
Without the presence of scalar field, Nojiri and Odintsov \cite{2}
presented $f(\mathcal{G})$ gravity as an alternative for DE by
adding generic function $f(\mathcal{G})$ in the Einstein-Hilbert
action. This theory elegantly describes the fascinating
characteristics of late-time cosmological evolution as well as
consistent with solar system constraints for a wide range of
cosmological viable model parameters \cite{3,3a}. Bamba et al.
\cite{d} investigated the finite-time future singularities and found
a possible way to cure these singularities in $f(\mathcal{G})$ as
well as $f(R,\mathcal{G})$ theories of gravity. Odintsov et al.
\cite {e} discussed the super-bounce and loop quantum ekpyrotic
cosmologies in the context of modified gravitational theories.

The non-minimal curvature-matter coupling in modified gravitational
theories has gained significant attention, since it can describe
consistently the late-time acceleration phenomenon. Harko et al.
\cite{4} proposed $f(R,T)$ theory of gravity as a generalization of
$f(R)$ gravity such that it involves the non-minimal coupling
between $R$ and $T$. Recently, we introduced such coupling in
$f(\mathcal{G})$ gravity referred as $f(\mathcal{G},T)$ theory and
found that the covariant divergence of energy-momentum tensor is not
zero \cite{5}. An extra force is appeared as a consequence of this
non-zero divergence due to which the non-geodesic trajectories are
followed by massive test particles while test particles with zero
pressure move along geodesic lines of geometry. The stability of
Einstein Universe against homogeneous isotropic and anisotropic
scalar perturbations is analyzed for both conserved as well as
non-conserved energy-momentum tensor in this theory and found stable
results \cite{6,6a}. Shamir and Ahmad \cite{7} constructed some
cosmological viable $f(\mathcal{G},T)$ model using Noether symmetry
approach in the context of homogeneous and isotropic Universe. The
background of cosmic evolutionary models corresponding to de Sitter
Universe, power-law solution as well as phantom/non-phantom epochs
can be reproduced in this theory \cite{8}.

Dynamical DE models have been constructed in the framework of
general relativity and quantum gravity which play an important role
to explore the mystery of cosmic expansion. Li \cite{9} proposed
holographic DE in the background of quantum gravity using the basic
concept of holographic principle which stands on the unified pillars
of quantum mechanics and gravity. This principle has gained much
importance by investigating quantum properties of black holes and
stimulated the attention of many researchers to explore string
theory or quantum gravity \cite{10}. Cohen et al. \cite{11}
reconciled Bekenstein's entropy bound by establishing a relationship
between ultraviolet and infrared cutoffs due to the limit made by
the black hole formation. In other words, the total energy of a
system with size $L$ should not be greater than the mass of black
hole with the same size for the quantum zero-point energy density
associated with the ultraviolet cutoff. This leads to the inequality
$L^{3}\rho_{_{\Lambda}}\leq LM_{_P}^{2}$, where $M_{_{P}}^{2}=(8\pi
G)^{-1}$ ($G$ is the gravitational constant), $L$ and
$\rho_{_{\Lambda}}$ are the reduced Planck mass, infrared cutoff and
energy density of holographic DE, respectively.

The accelerated expansion of the Universe is also successfully
discussed in literature via correspondence scheme of dynamical DE
models with modified theories of gravity. In this mechanism, generic
function of the considered gravity is reconstructed by comparing the
corresponding energy densities. A variety of reconstructed
holographic DE models in different modified theories have gained
remarkable importance in describing the present cosmic phase. Setare
\cite{12} examined the cosmological evolution of holographic DE in
$f(R)$ gravity for the flat FRW Universe model and found that the
reconstructed model behaves like phantom epoch of DE dominated era.
Setare and Saridakis \cite{13} developed a correspondence between
holographic DE scenario in flat FRW Universe and phantom DE model in
GB gravity coupled with a scalar field and found that this
correspondence consistently leads to the cosmic accelerated
expansion. Karami and Khaledian \cite{14} reconstructed the new
agegraphic as well as holographic DE $f(R)$ models for both ordinary
as well as entropy corrected version in flat FRW Universe model.
They found that both ordinary models behave like phantom or
non-phantom while the entropy corrected reconstructed models
experience the phase transition from quintessence to phantom epochs
of the Universe.

Houndjo and Piattella \cite{15} reconstructed holographic DE
$f(R,T)$ model numerically and observed that the same cosmic history
may be discussed by holographic DE model as in general relativity.
Daouda and his collaborators \cite{16} formulated the holographic DE
model in generalized teleparallel theory and concluded that the
resultant model implies unified mechanism of dark matter with DE.
Jawad et al. \cite{17} analyzed the stability of this dynamical DE
model with Granda-Oliveros cutoff in $f(\mathcal{G})$ gravity using
emergent, intermediate as well as logamediate scale factor and found
that the derived model is stable only for the intermediate case.
Sharif and Zubair \cite{18} investigated the holographic as well as
new agegraphic DE model in $f(R,T)$ gravity and observed that the
reconstructed models can demonstrate the phantom or quintessence
phases. They also discussed the generalized second law of
thermodynamics for the derived models and established the viability
conditions. Fayaz et al. \cite{19} found that the reconstructed
$f(R,T)$ models corresponding to holographic as well as new
ageagraphic DE in the context of Bianchi type I Universe model
illustrate phantom or quintessence regions.

In curvature-matter coupled gravitational theories, various
dynamical DE models have also gained significant importance in
describing the cosmic evolutionary phases. Sharif and Zubair
\cite{r1} considered the Ricci and modified Ricci DE models to
establish the equivalence between these dynamical DE models and
$f(R,T)$ gravity via reconstruction technique. They discussed the
Dolgov-Kawasaki instability criteria to explore the viability of
reconstructed models and found that the appropriate choice of
parameters in explaining the evolution of $f(R,T)$ models is
consistent with the viability conditions. Zubair and Abbas \cite{r2}
reconstructed the $f(R,T)$ theory for modified as well as
Garcia-Salcedo ghost DE models and analyzed the stability of
reconstructed ghost $f(R,T)$ models in the background of flat FRW
Universe model. They found that reconstructed ghost models elegantly
describe the phantom and quintessence regimes of the Universe. Fayaz
et al. \cite{r3} studied the anisotropic Universe with ghost DE
model and found that the reconstructed $f(R,T)$ models can reproduce
the cosmic phantom epoch satisfying the current observations. Baffou
and his collaborators \cite{ahep1} investigated the generalized
Chaplygin gas interacting with $f(R,T)$ theory of gravity in the
presence of bulk as well as shear viscosities and found that the
viscous parameters are well accommodated with observational data.
Zubair et al. \cite{ahep2} explored the cosmic evolution of $f(R,T)$
gravity in the presence of matter fluids consisting of radiation as
well as collisional self-interacting dark matter. Tiwari et al.
\cite{ahep3} constructed the cosmological model with variable
deceleration parameter in the background of $f(R,T)$ gravity. Sharif
and Saba \cite{ahep12} examined the pilgrim DE model and found that
the obtained $f(\mathcal{G},T)$ model illustrates the aggressive
phantom-like Universe as well as self-consistent pilgrim DE
$f(\mathcal{G},T)$ model.

In this paper, we investigate the cosmological evolution of
holographic DE in $f(\mathcal{G},T)$ gravity for flat FRW Universe
model. The paper has the following format. In section \textbf{2}, we
discuss basic formalism of this gravity, holographic DE and
reconstruct holographic DE $f(\mathcal{G},T)$ model using the
correspondence scheme. To its qualitative analysis, we consider
power-law form of the scale factor which may produce type III
finite-time future singularity. Section \textbf{3} is devoted to
examine the evolutionary behavior of equation of state (EoS) and
deceleration parameters as well as investigate the $r-s$ and
$\omega_{_{\mathcal{G}T}}-{\omega}'_{_{\mathcal{G}T}}$ cosmological
planes. The results are summarized in the last section.

\section{$f(\mathcal{G},T)$ Gravity and Holographic DE Model}

In this section, we briefly discuss basic concepts related to
$f(\mathcal{G},T)$ gravity and formulate the holographic DE
$f(\mathcal{G},T)$ model using the correspondence scenario. The
action for $f(\mathcal{G},T)$ theory of gravity is given by \cite{5}
\begin{equation}\label{1}
\mathcal{S}=\int
d^{4}x\sqrt{-g}\left(\frac{R+f(\mathcal{G},T)}{2\kappa^{2}}
+\mathcal{L}_{m}\right),
\end{equation}
where $g,~T=g_{\xi\eta}T^{\xi\eta},~\kappa^{2}$ and
$\mathcal{L}_{m}$ represent determinant of the metric tensor, trace
of the energy-momentum tensor, coupling constant and the Lagrangian
density associated with cosmic matter contents, respectively. The
variation of the above action with respect to $g_{\xi\eta}$ gives
the following fourth-order field equations
\begin{eqnarray}\nonumber
R_{\xi\eta}-\frac{1}{2}g_{\xi\eta}R&=&\kappa^{2}T_{\xi\eta}
+\frac{1}{2}g_{\xi\eta}f(\mathcal{G},T)-(T_{\xi\eta}+\Theta_{\xi\eta})
f_{T}(\mathcal{G},T)\\\nonumber&+&(4R_{\mu\eta}R^{\mu}_{\xi}
-2RR_{\xi\eta}+4R_{\xi\mu\eta\nu}R^{\mu\nu}-2R_{\xi}^{~\mu\nu\varrho}
R_{\eta\mu\nu\varrho})f_{\mathcal{G}}(\mathcal{G},T)\\\nonumber&+&
(4R_{\xi\eta}-2Rg_{\xi\eta})\nabla^{2}f_{\mathcal{G}}(\mathcal{G},T)
+2R\nabla_{\xi}\nabla_{\eta}f_{\mathcal{G}}(\mathcal{G},T)
\\\nonumber&-&4R^{\mu}_{\eta}\nabla_{\xi}\nabla_{\mu}f_{\mathcal{G}}
(\mathcal{G},T)-4R^{\mu}_{\xi}\nabla_{\eta}\nabla_{\mu}f_{\mathcal{G}}
(\mathcal{G},T)\\\label{2}&+&4g_{\xi\eta}R^{\mu\nu}\nabla_{\mu}
\nabla_{\nu}f_{\mathcal{G}}(\mathcal{G},T)-4R_{\xi\mu\eta\nu}
\nabla^{\mu}\nabla^{\nu}f_{\mathcal{G}}(\mathcal{G},T)=0,
\end{eqnarray}
where $\Theta_{\xi\eta}=g^{\mu\nu}(\delta T_{\mu\nu}/\delta
g^{\xi\eta})$ and $\nabla^{2}=\nabla_{\xi}\nabla^{\xi}$
($\nabla_{\xi}$ is the covariant derivative) whereas the subscripts
$T$ and $\mathcal{G}$ denote derivatives of generic function
$f(\mathcal{G},T)$ with respect to $T$ and $\mathcal{G}$,
respectively. The energy-momentum tensor is defined as \cite{20}
\begin{equation}\label{2a}
T_{\xi\eta}=-\frac{2}{\sqrt{-g}}\frac{\delta(\sqrt{-g}\mathcal{L}_{m})}{\delta
g^{\xi\eta}}.
\end{equation}
The covariant derivative of Eq.(\ref{2}) gives
\begin{eqnarray}\nonumber
\nabla^{\xi}T_{\xi\eta}&=&\frac{f_{T}(\mathcal{G},T)}{\kappa^{2}-f_{T}
(\mathcal{G},T)}\left[(\Theta_{\xi\eta}+T_{\xi\eta})\nabla^{\xi}\ln
f_{T}(\mathcal{G},T)-\frac{1}{2}g_{\xi\eta}
\nabla^{\xi}T\right.\\\label{5}&+&\left.\nabla^{\xi}\Theta_{\xi\eta}\right].
\end{eqnarray}
This shows that the energy-momentum tensor is not conserved due to
the coupling present between geometry and matter contents.

In curvature-matter coupled theories, the generic function and
matter Lagrangian density play a pivotal role to explore their
dynamics. Some particular forms are as follows.
\begin{itemize}
\item $f(\mathcal{G},T)=f_{1}(\mathcal{G})+f_{2}(T)$:
This choice is considered as correction to $f(\mathcal{G})$ theory
of gravity since direct non-minimal curvature-matter coupling is
absent. It is remarked that $f_{1}(\mathcal{G})$ cannot be linearly
taken as GB invariant ($f_{1}(\mathcal{G})\neq\mathcal{G}$) due to
its four-dimensional topological behavior.
\item
$f(\mathcal{G},T)=f_{1}(\mathcal{G})+f_{2}(\mathcal{G})f_{3}(T)$:
This model involves direct non-minimal coupling whose consequences
would be different from the above form of generic function.
\end{itemize}
The line element for flat FRW Universe model is
\begin{equation}\label{4}
ds^2=dt^2-a^2(t)(dx^2+dy^2+dz^2),
\end{equation}
where $a(t)$ is the scale factor depending on cosmic time $t$. We
consider perfect fluid configuration as cosmic matter content with
$\mathcal{L}_{m}=-p$. The corresponding energy-momentum tensor and
$\Theta_{\xi\eta}$ are given by
\begin{equation}\label{5}
T_{\xi\eta}=(\rho+p)V_{\xi}V_{\eta}-pg_{\xi\eta},\quad
\Theta_{\xi\eta}=-2T_{\xi\eta}-pg_{\xi\eta},
\end{equation}
where $\rho,~p$ and $V_{\xi}$ are the energy density, pressure and
four-velocity of the fluid, respectively. The expressions for trace
of energy-momentum tensor and GB invariant are
\begin{equation}\label{5a}
T=\rho-3p,\quad\mathcal{G}=24H^{2}(\dot{H}+H^{2}),
\end{equation}
where $H=\dot{a}/a$ is the Hubble parameter and dot represents time
derivative. In this analysis, we have considered the units in which
$\kappa^{2}=1$. Using Eqs.(\ref{4}) and (\ref{5}) in (\ref{2}), we
obtain the following field equations
\begin{equation}\label{6}
3H^2=\rho+\rho_{_{\mathcal{G}T}},\quad
-(2\dot{H}+3H^2)=p+p_{_{\mathcal{G}T}},
\end{equation}
where \begin{eqnarray}\nonumber
\rho_{_{\mathcal{G}T}}&=&\frac{1}{2}f(\mathcal{G},T)+(\rho+p)f_{_{T}}
(\mathcal{G},T)-12H^{2}(\dot{H}+H^{2})f_{_{\mathcal{G}}}(\mathcal{G},T)
\\\nonumber&+&12H^{3}(f_{_{\mathcal{GG}}}(\mathcal{G},T)\dot{\mathcal{G}}
+f_{_{\mathcal{G}T}}(\mathcal{G},T)\dot{T}),\\\nonumber
p_{_{\mathcal{G}T}}&=&-\frac{1}{2}f(\mathcal{G},T)+12H^{2}(\dot{H}+H^{2})
f_{_{\mathcal{G}}}(\mathcal{G},T)-8H(\dot{H}+H^{2})(f_{_{\mathcal{GG}}}
(\mathcal{G},T)\dot{\mathcal{G}}\\\nonumber&+&f_{_{\mathcal{G}T}}(\mathcal{G},T)
\dot{T})-4H^{2}(f_{_{\mathcal{GG}}}(\mathcal{G},T)\ddot{\mathcal{G}}
+f_{_{\mathcal{G}T}}(\mathcal{G},T)\ddot{T}+2f_{_{\mathcal{GG}T}}(\mathcal{G},T)
\dot{\mathcal{G}}\dot{T}\\\nonumber&+&f_{_{\mathcal{GGG}}}(\mathcal{G},T)
\dot{\mathcal{G}}^{2}+f_{_{\mathcal{G}TT}}(\mathcal{G},T)\ddot{T}^{2}).
\end{eqnarray}

The energy density of holographic DE model is given by \cite{12}
\begin{equation}\label{9}
\rho_{_\Lambda}=\frac{3\tilde{c}^{2}}{\mathcal{R}_{h}^{2}},
\end{equation}
where $\mathcal{R}_{h}^{2}$ denotes the future event horizon
(infrared cutoff) defined as \cite{9}
\begin{equation}\label{10}
\mathcal{R}_{h}^{2}=a\int_{a}^{\infty}\frac{da}{a^{2}H}
=a\int^{\infty}_{t}\frac{dt}{a}.
\end{equation}
Differentiating this relation with respect to time, we obtain
\begin{equation}\label{11}
\dot{\mathcal{R}}_{h}=H\mathcal{R}_{h}-1=\frac{\tilde{c}}
{\sqrt{\Omega_{_{\Lambda}}}}-1,
\end{equation}
where $\Omega_{_{\Lambda}}$ is the ratio between holographic and
critical $(\rho_{_{c}}=3H^2)$ energy densities dubbed as
dimensionless DE. The EoS parameter for this DE model is given by
\begin{equation}\label{12}
\omega_{_{\Lambda}}=-\frac{1}{3}\left(\frac{2\sqrt{\Omega_{_{\Lambda}
}}}{\tilde{c}}+1\right).
\end{equation}
At the early times with $\Omega_{_{\Lambda}}\ll1$, the holographic
DE subdominants the cosmic contents leading to $\omega\approx-1/3$
while it dominates at the late Universe with
$\Omega_{_{\Lambda}}\approx1$. In this case, the behavior of
$\omega_{_{\Lambda}}$ depends on the values of parameter
$\tilde{c}$. The holographic DE represents the phantom
($\omega_{_{\Lambda}}<-1$) and non-phantom
$(\omega_{_{\Lambda}}>-1)$ phases of the Universe for $\tilde{c}<1$
and $\tilde{c}>1$, respectively while it demonstrates the de Sitter
Universe $(\omega_{\Lambda}=-1)$ for $\tilde{c}=1$. Thus, the
parameter $\tilde{c}$ plays a pivotal role in determining the cosmic
evolutionary scenario of holographic DE. It is worth mentioning here
that its value cannot be obtained from any theoretical framework
rather than it has been constrained only from observational data. In
our analysis, we choose the best fitted value $\tilde{c}=0.506$ at
the $68\%$C.L. (C.L. stands for confidence level) constrained from
observational data of Planck+WP+BAO (WP and BAO are Wilkinson
microwave anisotropy probe 9 polarization data and baryon acoustic
oscillations, respectively) which favors the phantom behavior of
holographic DE model \cite{21}.

Now we reconstruct the holographic DE $f(\mathcal{G},T)$ model using
the paradigm of correspondence scheme. For the sake of simplicity,
we consider pressureless fluid configuration with the particular
form of $f(\mathcal{G},T)$ as \cite{6}
\begin{equation}\label{13}
f(\mathcal{G},T)=F(\mathcal{G})+\chi T,
\end{equation}
where $\chi$ is an arbitrary constant. The field equations for this
choice of generic function reduce to
\begin{equation}\label{14}
3H^2=\rho+\rho_{_{\mathcal{G}T}},\quad
-(2\dot{H}+3H^2)=p_{_{\mathcal{G}T}},
\end{equation}
where
\begin{eqnarray}\label{15}
\rho_{_{\mathcal{G}T}}&=&\frac{3}{2}\chi\rho+\frac{1}{2}F(\mathcal{G})
-12H^{2}(\dot{H}+H^{2})F'(\mathcal{G})+12H^{3}\dot{\mathcal{G}}
F''(\mathcal{G}),\\\nonumber p_{_{\mathcal{G}T}}&=&
-\frac{1}{2}\chi\rho-\frac{1}{2}F(\mathcal{G})+12H^{2}(\dot{H}+H^{2})
F'(\mathcal{G})-8H(\dot{H}+H^{2})\dot{\mathcal{G}}F''(\mathcal{G})\\\label{16}
&-&4H^{2}[\ddot{\mathcal{G}}F''(\mathcal{G})+\dot{\mathcal{G}}^{2}
F'''(\mathcal{G})],
\end{eqnarray}
$\kappa^{2}=1$ and prime represents derivative with respect to
$\mathcal{G}$. The addition of Eqs.(\ref{15}) and (\ref{16}) yields
the third order non-linear differential equation in $F(\mathcal{G})$
as follows
\begin{eqnarray}\label{17}
\chi\rho-(\rho_{_{\mathcal{G}T}}+p_{_{\mathcal{G}T}})+4H[(H^2-2\dot{H})
\dot{\mathcal{G}}-H\ddot{\mathcal{G}}]F''(\mathcal{G})-4H^2\dot{\mathcal{G}}^2
F'''(\mathcal{G})=0.
\end{eqnarray}

In order to obtain its solution, we consider power-law form of the
scale factor which has a significant importance in cosmology since
it elegantly illustrates different cosmic evolutionary phases given
by \cite{12}
\begin{equation}\label{18}
a(t)=a_{\circ}(\tau-t)^{\lambda},\quad\tau>t,\quad\lambda>0,
\end{equation}
where $a_{\circ}$ and $\tau$ represent the present day value of
$a(t)$ and finite-time future singularity, respectively. The
accelerated phase of the Universe is observed for $\lambda>1$
whereas $0<\lambda<1$ covers the decelerated phase including dust
($\lambda=2/3$) as well as radiation ($\lambda=1/2$) dominated
epoch. The finite-time future singularities are the timelike
singularities which are classified into four types depending on
physical quantities ($a(t)$, effective pressure
$(p_{_{\mathrm{eff}}}=p+p_{_{\mathcal{G}T}})$ and effective energy
density ($\rho_{_{\mathrm{eff}}}=\rho+\rho_{_{\mathcal{G}T}}$))
\cite{f}. The big-rip singularity is usually referred as type I
singularity in which all these physical variables diverge as
$t\rightarrow\tau$ while in type II singularity, only effective
pressure diverges as cosmic time approaches $\tau$. In case of type
III singularity, $a(t)$ remains finite while the total energy
density and pressure diverge as $t\rightarrow\tau$. For type IV
finite-time singularity, all physical quantities as well as Hubble
rate along with its first derivative are finite as
$t\rightarrow\tau$ while higher derivatives diverge. A bounce
cosmology with type IV singularity at a bouncing point is also
investigated in the context of $f(\mathcal{G})$ gravity
\cite{ahep8}. These singularities in the context of various
gravitational theories are studied in literature
\cite{d,ahep4}-\cite{ahep7}. Using Eq.(\ref{18}), the expressions of
$H,~\mathcal{G},~\mathcal{R}_{h},~\rho_{_{\Lambda}},~\rho$ and
$p_{_{\mathcal{G}T}}$ take the form
\begin{eqnarray}\label{19}
&&H=-\frac{\lambda}{\tau-t},\quad\mathcal{G}=\frac{24\lambda^{3}(\lambda-1)}
{(\tau-t)^4},\quad\mathcal{R}_{h}=\frac{\tau-t}{1-\lambda},\\\label{20}&&
\rho_{_{\Lambda}}=\frac{3\tilde{c}^{2}(1-\lambda)^{2}}{(\tau-t)^2},\quad\rho
=\frac{3[\lambda^{2}-\tilde{c}^2(1-\lambda)^{2}]}{(\tau-t)^{2}},\quad
p_{_{\mathcal{G}T}} =\frac{\lambda(2-3\lambda^{2})}{(\tau-t)^2}.
\end{eqnarray}
According to the above functional forms of the effective energy
density and pressure, this cosmological evolution leads to a type
III singularity at $t=\tau$. This is also obvious from the
functional form of the Hubble rate which diverges at $t=\tau$. By
applying the correspondence of energy densities
$(\rho_{_{\Lambda}}=\rho_{_{\mathcal{G}T}})$ and substituting
Eqs.(\ref{19}) and (\ref{20}) in (\ref{17}), the resultant
differential equation becomes
\begin{equation}\label{21}
\Delta_{1}\mathcal{G}^{\frac{1}{2}}+\Delta_{2}\mathcal{G}^{2}F''(\mathcal{G})
+\Delta_{3}\mathcal{G}^{3}F'''(\mathcal{G})=0,
\end{equation}
where
\begin{eqnarray}\nonumber
\Delta_{1}&=&\frac{3\chi[\lambda^{2}-\tilde{c}^2(1-\lambda)^{2}]-3\tilde{c}^{2}(1-\lambda)^{2}
-\lambda(2-3\lambda^{2})}{[24\lambda^{3}(\lambda-1)]^{\frac{1}{2}}},\\\nonumber
\Delta_{2}&=&-\frac{2(\lambda+7)}{3\lambda(\lambda-1)},\quad\Delta_{3}=-\frac{8}
{3\lambda(\lambda-1)}.
\end{eqnarray}
Its solution is given by
\begin{equation}\label{22}
F(\mathcal{G})=d_{1}+d_{2}\mathcal{G}-\frac{d_{3}\mathcal{G}^{2-\frac{\Delta_{2}}
{\Delta_{3}}}}{\left(2-\frac{\Delta_{2}}{\Delta_{3}}\right)(\Delta_{2}-\Delta_{3})}
+\frac{8\Delta_{1}\sqrt{\mathcal{G}}}{2\Delta_{2}-3\Delta_{3}},
\end{equation}
where $d_{i}$'s $(i=1,2,3)$ are the integration constants.
Consequently, the reconstructed $f(\mathcal{G},T)$ model
corresponding to holographic DE is
\begin{equation}\label{23}
f(\mathcal{G},T)=d_{1}+d_{2}\mathcal{G}-\frac{d_{3}\mathcal{G}^{2-\frac{\Delta_{2}}
{\Delta_{3}}}}{\left(2-\frac{\Delta_{2}}{\Delta_{3}}\right)(\Delta_{2}-\Delta_{3})}
+\frac{8\Delta_{1}\sqrt{\mathcal{G}}}{2\Delta_{2}-3\Delta_{3}}+\chi
T,
\end{equation}
\begin{figure}
\epsfig{file=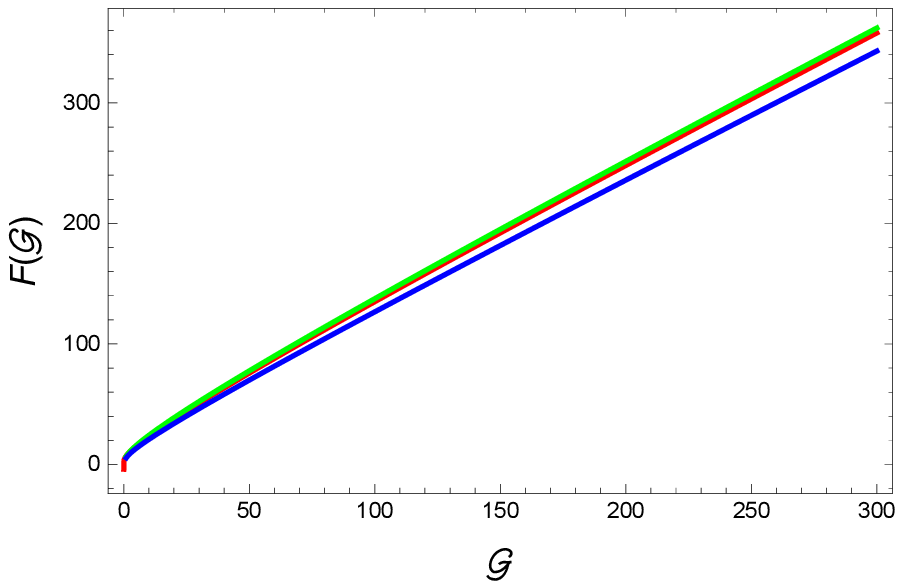, width=0.5\linewidth}\epsfig{file=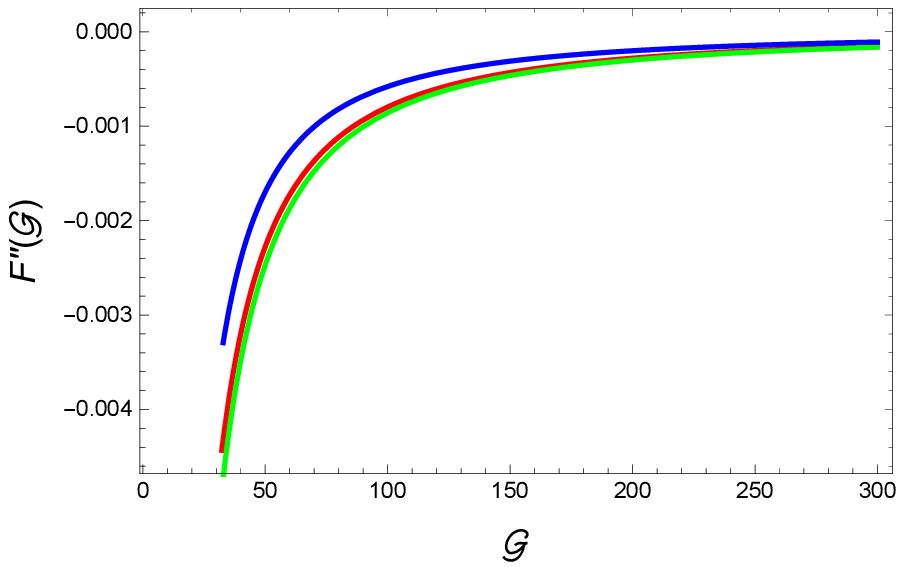,
width=0.5\linewidth}\caption{Evolution of holographic DE
$F(\mathcal{G})$ model (left) and $F''(\mathcal{G})$ (right) versus
$\mathcal{G}$ for $\lambda=1.5$ (red), $\lambda=2$ (green) and
$\lambda=2.4$ (blue).}
\end{figure}

Figure \textbf{1} shows the graphical analysis of holographic DE
$F(\mathcal{G})$ model in the left panel while the right plot
demonstrates its stability with the parameters chosen as
$d_{1}=5,~d_{2}=d_{3}=1$ and $\chi=-2.74$ throughout the analysis.
We observe that the reconstructed model exhibits positively
increasing behavior as $\mathcal{G}$ increases while it approaches
zero as $\mathcal{G}\rightarrow0$ for all the considered values of
$\lambda$. It is important to mention here that stability of any
$F(\mathcal{G})$ model depends on the regularity of generic function
and its derivatives along with the condition $F''(\mathcal{G})<0$
for metric signatures $(+,-,-,-)$ while reverse inequality is
required for the second choice of signatures for all $\mathcal{G}$
\cite{22,22a}. Thus, the right plot shows that stability condition
is satisfied for the reconstructed holographic DE $F(\mathcal{G})$
model.

\section{Cosmological Analysis}

In this section, we analyze the EoS and deceleration parameters as
well as examine the cosmological planes such as $r-s$ and
$\omega_{_{\mathcal{G}T}}-{\omega}'_{_{\mathcal{G}T}}$ for the
reconstructed holographic DE $f(\mathcal{G},T)$ model.

\subsection{EoS Parameter}

The EoS parameter for the obtained model using the correspondence
scenario of energy densities is given by
\begin{equation}\label{eos}
\omega_{_{\mathcal{G}T}}=\frac{p_{_{\mathcal{G}T}}}{\rho_{\mathcal{G}T}}
=\frac{p_{_{\mathcal{G}T}}}{\rho_{_{\Lambda}}}.
\end{equation}
Carroll et al. \cite{eos1} found that any phantom model with EoS
parameter less than $-1$ should decay to $\omega=-1$ at late time in
the context of general relativity using the scalar field Lagrangian
density. Amirhashchi \cite{eos2} observed that presence of bulk
viscosity in the cosmic fluid can temporarily drive the fluid into
the phantom region and ultimately EoS parameter of DE approaches to
$-1$ as time passes. The presence of bulk viscosity in the
background of anisotropic Bianchi I line element causes transition
of EoS parameter of DE from quintessence to phantom which also
decays to $-1$ at late time \cite{eos3}. Amirhashchi \cite{eos4}
also analyzed the behavior of DE and found a possibility of DE EoS
parameter to cross the phantom divide line for anisotropic Bianchi V
spacetime.

We use scale factor in terms of red-shift parameter as
$a=a_{\circ}(1+z)^{-1}$ throughout the graphical analysis. Figure
\textbf{2} shows the cosmic evolutionary picture of EoS parameter
against red-shift parameter $(z)$ using holographic DE model for
$\lambda=2.4$. It is observed that EoS parameter represents the
phantom regime at present $(z=0)$ and the corresponding value is
$\omega_{_{\mathcal{G}T}}=-1.2$ consistent with Planck observational
data \cite{ahep9} as well as in agreement with tilted flat and
untitled non-flat XCDM model parameters constrained from Planck data
\cite{ahep10}. It is also demonstrated from the graphical analysis
that this parameter remains in the phantom regime and may not have a
possibility to decay to $\omega_{_{\mathcal{G}T}}=-1$ at late time.
Here, the phantom phase of the Universe is consistent with the
observational data of holographic DE parameter $\tilde{c}$.
\begin{figure}\centering
\epsfig{file=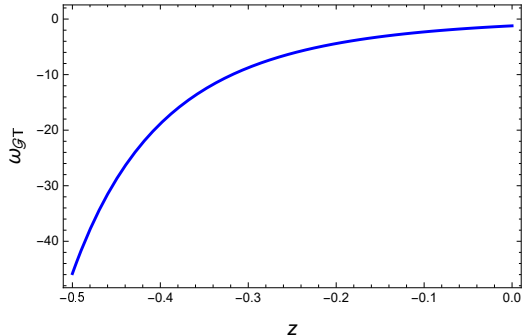, width=0.5\linewidth}\caption{Evolution of
EoS parameter versus $z$ for $\lambda=2.4$.}
\end{figure}

\subsection{Deceleration Parameter}

The deceleration parameter is defined as
\begin{equation}\label{dec}
q=-\frac{a\ddot{a}}{\dot{a}^{2}}=-1-\frac{\dot{H}}{H^2}.
\end{equation}
Its positive value indicates the cosmic decelerated phase while
negative value characterizes the epoch of accelerated expansion.
Figure \textbf{3} shows the graphical cosmological evolution of
deceleration parameter for the reconstructed model (\ref{23})
against $z$. We observe that the value of this parameter is $-0.53$
at $z=0$ which is consistent with observational data of Planck
\cite{ahep9} as well as favors the current constraints on isotropic
and anisotropic DE models \cite{ahep11}. Thus, the holographic DE
$f(\mathcal{G},T)$ model demonstrates the accelerating phase of the
cosmic expansion.
\begin{figure}\centering
\epsfig{file=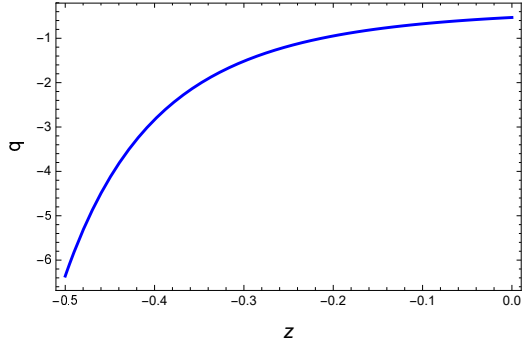, width=0.5\linewidth}\caption{Evolution of
deceleration parameter versus $z$ for $\lambda=2.4$.}
\end{figure}

\subsection{$r-s$ Plane}

Sahni et al. \cite{23} introduced the cosmological diagnostic pair
of dimensionless parameters known as state-finder diagnostic
parameters to discriminate DE models such that one can determine
which model is more suitable for a better explanation of the current
cosmic status. These parameters are defined as
\begin{equation}\label{rs}
r=\frac{\dddot{a}}{aH^{3}}=2q^{2}+q-\frac{\dot{q}}{H},\quad
s=\frac{r-1}{3\left(q-\frac{1}{2}\right)}.
\end{equation}
\begin{figure}\centering
\epsfig{file=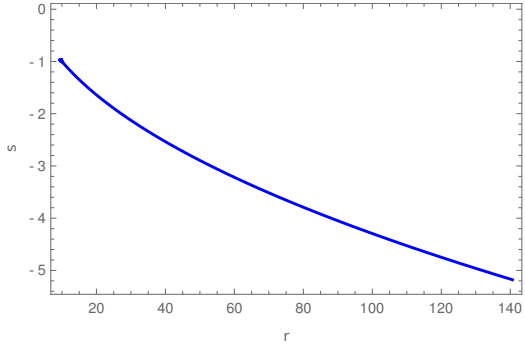, width=0.5\linewidth}\caption{Plot of
state-finder diagnostic parameters for $\lambda=2.4$.}
\end{figure}
The plane of these cosmological parameters (dubbed as $r-s$ plane)
for $\Lambda$CDM model (CDM stands for cold dark matter) is fixed as
$(r,s)=(1,0)$ while $(r,s)=(1,1)$ corresponds to CDM regime. The
phantom as well as non-phantom DE epochs are illustrated by the
regions ($r<1,s>0$) whereas trajectories for Chaplygin gas lie in
the range ($r>1,s<0$). Figure \textbf{4} shows graphical
interpretation of holographic DE $f(\mathcal{G},T)$ model in $r-s$
plane and observed that the evolutionary trajectory only correspond
to the Chaplygin gas model.

\subsection{$\omega_{_{\mathcal{G}T}}-{\omega}'_{_{\mathcal{G}T}}$ Plane}
\begin{figure}\centering
\epsfig{file=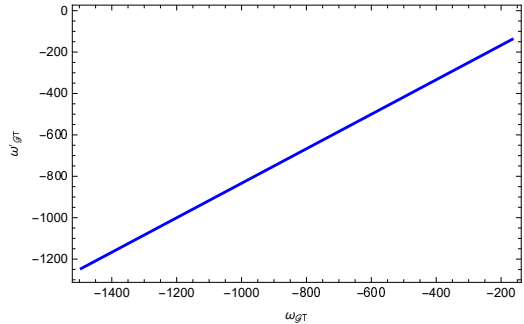, width=0.5\linewidth}\caption{Plot of
$\omega_{_{\mathcal{G}T}}-{\omega}'_{_{\mathcal{G}T}}$ for
$\lambda=2.4$.}
\end{figure}

Caldwell and Linder \cite{24} presented
$\omega_{_{\mathcal{G}T}}-{\omega}'_{_{\mathcal{G}T}}$
(${\omega}'_{_{\mathcal{G}T}}$ is the evolutionary form of
${\omega}_{_{\mathcal{G}T}}$ defined as
${\omega}'_{_{\mathcal{G}T}}=d{\omega}_{_{\mathcal{G}T}}/d\ln a$)
plane to investigate cosmic evolution of quintessence scalar field
DE model and found that area occupied by the considered model in
this plane can be categorized into freezing $(\omega<0,~\omega'<0)$
and thawing ($\omega<0,~\omega'>0$) regions. It is remarked that
cosmic expansion is more accelerating in freezing region as compared
to thawing. The graphical interpretation of
$\omega_{_{\mathcal{G}T}}-{\omega}'_{_{\mathcal{G}T}}$ is shown in
Figure \textbf{5} which indicates the freezing region.

\section{Final Remarks}

In this paper, we have explored cosmological reconstruction of
$f(\mathcal{G},T)$ gravity with a well-known holographic DE model
using the power-law scale factor. The accelerated expansion of the
Universe is considered as an outcome of integrated contribution from
geometric and matter components. We have considered flat FRW
Universe with pressureless matter contribution as cosmic fluid
configuration and constructed the corresponding field equations for
a particular form $f(\mathcal{G},T)=F(\mathcal{G})+\chi T$ ($\chi$
is an arbitrary constant). To reconstruct the holographic DE
$F(\mathcal{G})$ model, we have applied the correspondence scheme by
comparing the corresponding energy densities. The derived model
possesses increasing behavior as well as satisfies the stability
condition (Figure \textbf{1}).

We have examined the evolutionary paradigm of reconstructed
holographic $f(\mathcal{G},T)$ model through EoS and deceleration
parameters as well as $r-s$ and
$\omega_{_{\mathcal{G}T}}-{\omega}'_{_{\mathcal{G}T}}$ cosmological
planes. The results are summarized as follows.
\begin{itemize}
\item The trajectory of EoS parameter indicates the phantom phase of
the Universe for the considered value of $\lambda$ at $z=0$ (Figure
\textbf{2}). \item The evolution of deceleration parameter against
cosmic time gives accelerated phase of the Universe throughout the
evolution (Figure \textbf{3}). \item The state-finder diagnostic
plane for the reconstructed model only corresponds to Chaplygin gas
model (Figure \textbf{4}). \item The trajectory in
$\omega_{_{\mathcal{G}T}} -{\omega}'_{_{\mathcal{G}T}}$ plane
represents the freezing regime for the considered value of
$\lambda$. Hence,
$\omega_{_{\mathcal{G}T}}-{\omega}'_{_{\mathcal{G}T}}$ plane shows
consistency with the cosmic accelerated expansion (Figure
\textbf{5}).
\end{itemize}

\textbf{Acknowledgement}\\

We would like to thank the Higher Education Commission, Islamabad,
Pakistan for its financial support through the \emph{Indigenous
Ph.D. Fellowship for 5000 Scholars, Phase-II, Batch-III}.


\begin{thebibliography}{30}

\bibitem{a} Nojiri, S. and Odintsov, S.D.: Phys. Rev. D
\textbf{68}(2003)123512.

\bibitem{b} Nojiri, S. and Odintsov, S.D.: Int. J. Geom. Methods Mod. Phys.
\textbf{04}(2007)115.

\bibitem{b1} Nojiri, S. and Odintsov, S.D.: Phys. Rept. \textbf{505}(2011)59.

\bibitem{b1a} Capozziello, S. and De Laurentis, M.: Phys. Rept.
\textbf{509}(2011)167.

\bibitem{b2} Bamba, K. et al.: Astrophys. Space
Sci. \textbf{342}(2012)155.

\bibitem{b3} Nojiri, S., Odintsov, S.D. and Oikonomou, V.K.: Phys. Rept.
\textbf{692}(2017)1.

\bibitem{1} Bhawal, B. and Kar, S.: Phys. Rev. D \textbf{46}(1992)2464.

\bibitem{1a} Deruelle, N. and Dole$\check{z}$el, T.: Phys. Rev. D
\textbf{62}(2000)103502.

\bibitem{1b} de Felice, A. and Tsujikawa, S.: Living Rev. Rel.
\textbf{13}(2010)3.

\bibitem{c} Nojiri, S., Odintsov, S.D. and Sasaki, M.: Phys. Rev. D \textbf{71}(2005)123509.

\bibitem{2} Nojiri, S. and Odintsov, S.D.: Phys. Lett. B \textbf{631}(2005)1.

\bibitem{3} Cognola, G. et al.: Phys. Rev. D
\textbf{73}(2006)084007.

\bibitem{3a} de Felice, A. and Tsujikawa, S.: Phys. Rev. D
\textbf{80}(2009)063516.

\bibitem{d} Bamba, K. et al.: Eur. Phys. J. C \textbf{67}(2010)295.

\bibitem{e} Odintsov, S.D., Oikonomou, V.K. and Saridakis, E.N.: Ann. Phys. \textbf{363}(2015)141.

\bibitem{4} Harko, T. et al.: Phys. Rev. D \textbf{84}(2011)024020.

\bibitem{5} Sharif, M. and Ikram, A.: Eur. Phys. J. C \textbf{76}(2016)640.

\bibitem{6} Sharif, M. and Ikram, A.: Int. J. Mod. Phys. D \textbf{26}(2017)1750084.

\bibitem{6a} Sharif, M. and Ikram, A.: Eur. Phys. J. Plus \textbf{132}(2017)526.

\bibitem{7} Shamir, M.F. and Ahmad, M.: Eur. Phys. J. C \textbf{77}(2017)55.

\bibitem{8} Sharif, M. and Ikram, A.: Phys. Dark Universe \textbf{17}(2017)1.

\bibitem{9} Li, M.: Phys. Lett. B \textbf{603}(2004)1.

\bibitem{10} Susskind, L.: J. Math. Phys. \textbf{36}(1995)6377; Hooft,G.'t.: arXiv:gr-qc/9310026.

\bibitem{11} Cohen, A.G., Kaplan, D.B. and Nelson, A.E.: Phys. Rev. Lett.
\textbf{82}(1999)4971.

\bibitem{12} Setare, M.R.: Int. J. Mod. Phys. D \textbf{17}(2008)2219.

\bibitem{13} Setare, M.R. and Saridakis, E.N.: Phys. Lett. B \textbf{670}(2008)1.

\bibitem{14} Karami, K. and Khaledian, M.S.: J. High Energy Phys. \textbf{03}(2011)086.

\bibitem{15} Houndjo, M.J.S. and Piattella, O.F.: Int. J. Mod. Phys. D \textbf{21}(2012)1250024.

\bibitem{16} Daouda, M.H., Rodrigues, M.E. and Houndjo, M.J.S.: Eur. Phys. J. C \textbf{72}(2012)1893.

\bibitem{17} Jawad, A., Pasqua, A. and Chattopadhyay, S.: Eur. Phys. J. Plus \textbf{128}(2013)156.

\bibitem{18} Sharif, M. and Zubair, M.: J. Phys. Soc. Jpn. \textbf{82}(2013)064001.

\bibitem{19} Fayaz, V. et al.: Astrophys. Space Sci. \textbf{353}(2014)301.

\bibitem{r1} Sharif, M. and Zubair, M.: Astrophys. Space Sci \textbf{349}(2014)529.

\bibitem{r2} Zubair, M. and Abbas, G.: Astrophys. Space Sci.
\textbf{357}(2015)154.

\bibitem{r3} Fayaz, V. et al.: Eur. Phys. J. Plus
\textbf{131}(2016)22.

\bibitem{ahep1} Baffou, E.H., Houndjo, M.J.S. and Salako, I.G.: Int. J.
Geom. Methods Mod. Phys. \textbf{14}(2017)1750051.

\bibitem{ahep2} Zubair, M. et al.: Symmetry \textbf{10}(2018)463.

\bibitem{ahep3} Tiwari, R.K., Beesham, A. and Shukla, B.: Int. J. Geom.
Methods Mod. Phys. \textbf{15}(2018)1850115.

\bibitem{ahep12} Sharif, M. and Saba, S.: Mod. Phys. Lett. A
\textbf{33}(2018)1850182.

\bibitem{20} Landau, L.D. and Lifshitz, E.M.: \emph{The Classical Theory of Fields}
(Pergamon Press, 1971).

\bibitem{21} Li, M. et al.: J. Cosmol. Astropart. Phys. \textbf{09}(2013)021.

\bibitem{f} Nojiri, S., Odintsov, S.D. and Tsujikawa, S.:
Phys. Rev. D \textbf{71}(2005)063004.

\bibitem{ahep8} Oikonomou, V.K.: Phys. Rev. D \textbf{92}(2015)124027.

\bibitem{ahep4} Nojiri, S., Odintsov, S.D. and Tsujikawa, S.:
Phys. Rev. D \textbf{71}(2005)063004.

\bibitem{g} Nojiri, S., Odintsov, S.D. and Oikonomou, V.K.: Phys. Rev. D
\textbf{91}(2015)084059.

\bibitem{g1} Nojiri, S. et al.: J. Cosmol. Astropart.
Phys. \textbf{09}(2015)044.

\bibitem{g1a} Odintsov, S.D. and Oikonomou, V.K.: Phys. Rev. D
\textbf{92}(2015)024016.

\bibitem{g2} Oikonomou, V.K.: Int. J. Geom. Methods Mod. Phys.
\textbf{13}(2016)1650033.

\bibitem{ahep5} Bahamonde, S. et al.: Ann. Phys. \textbf{373}(2016)96.

\bibitem{ahep6} Odintsov, S.D. and Oikonomou, V.K.: Phys. Rev. D
\textbf{97}(2018) 124042.

\bibitem{ahep7} Odintsov, S.D. and Oikonomou, V.K.: Phys. Rev. D
\textbf{98}(2018)024013.

\bibitem{22} de Felice, A. and Tsujikawa, S.: Phys. Lett. B
\textbf{675}(2009)1.

\bibitem{22a} Li, B., Barrow, J.D. and Mota, D.F.: Phys. Rev. D
\textbf{76}(2007)044027.

\bibitem{eos1} Carroll, S.M., Hoffman, M. and Trodden, M.: Phys.
Rev. D \textbf{68}(2003)023509.

\bibitem{eos2} Amirhashchi, H.: Astrophys. Space Sci. \textbf{345}(2013)439.

\bibitem{eos3} Amirhashchi, H. and Pradhan, A.: Astrophys. Space Sci. \textbf{351}(2014)59.

\bibitem{eos4} Amirhashchi, H.: Phys. Rev. D \textbf{96}(2017)123507.

\bibitem{ahep9} Ade, P.A.R. et al.: Astron. Astrophys.:
\textbf{594}(2016)A13.

\bibitem{ahep10} Park, C.G. and Ratra, B.: arXiv:1803.05522.

\bibitem{ahep11} Amirhashchi, H. and Amirhashchi, S.: Phys. Rev. D
\textbf{99}(2019)023516.

\bibitem{23} Sahni, V. et al.: J. Exp. Theor. Phys. Lett. \textbf{77}(2003)201.

\bibitem{24} Caldwell, R.R. and Linder, E.V.: Phys. Rev. Lett. \textbf{95}(2005)141301.

\end{thebibliography}
\end{document}